\documentclass[11pt]{article}
\usepackage{mathtools} 
\usepackage{amssymb}    
\usepackage{amsmath}    
\usepackage{amsthm} 
\usepackage{color}  
\usepackage{enumitem}

\newtheorem{theorem}{Theorem}

\newtheorem{cor}{Corollary}
\newtheorem{lemma}{Lemma}
\newtheorem{defin}{Definition}

\newtheorem{condition}{Condition}

\def\Q{Q}
\def\R{{\mathbb R}}

\def\P{\mathcal{P}}

\def\U{\mathcal{U}}
\def\V{\mathcal{V}}
\def\Y{\mathcal{Y}}

\def\L{\mathcal{L}}
\def\PROB {{\mathbb P}}
\def\EXP {{\mathbb E}}
\def\IND{{\mathbb I}}

\def\argmax{\mathop{\rm arg\,max}}

\newcommand\ip[2]{ \left\langle{#1} \, ,\, {#2} \right\rangle } 

\begin{document}
\thispagestyle{empty}
\setcounter{page}{1}

\title{Lossless Transformations and Excess Risk Bounds  \\in Statistical Inference}
\author{
L\'aszl\'o Gy\"orfi\thanks{Budapest University of Technology and Economics, gyorfi@cs.bme.hu. The research of L\'aszl\'o Gy\"orfi has been supported by the National Research, Development and Innovation Fund of Hungary under the 2019-1.1.1-PIACI-KFI-2019-00018 funding
scheme.}
\and
Tam\'as Linder\thanks{Queen's University, Kingston, ON Canada, tamas.linder@queensu.ca. Tam\'as Linder's research was supported in part by the Natural Sciences and Engineering Research Council (NSERC) of Canada.}
\and
Harro Walk\thanks{Universit\"at Stuttgart, harro.walk@mathematik.uni-stuttgart.de} }

\maketitle

\begin{abstract}
We study the excess minimum risk in statistical inference, defined as the difference between the minimum expected loss in estimating a random variable from an observed  feature vector and the minimum expected loss in estimating the same random variable from a transformation (statistic) of the feature vector. After characterizing lossless transformations, i.e., transformations for which the excess risk is zero for all loss functions, we construct a partitioning test statistic 
for the hypothesis that a given transformation is lossless and show that for i.i.d.\ data the test is strongly consistent. More generally, we develop information-theoretic upper bounds on the excess risk that uniformly hold over  fairly general classes of loss functions. Based on these bounds, we introduce the notion of a $\delta$-lossless transformation and give sufficient conditions for a given transformation to be universally $\delta$-lossless. Applications  to classification, nonparametric regression, portfolio strategies, information bottleneck,  and deep learning, are also surveyed. 
\end{abstract}

\noindent


\noindent




\section{Introduction}

\label{sec-int}

We consider the standard setting of statistical inference, where $Y$ is a real random variable, having range $\Y\subset \R$, that is to be estimated (predicted) from a random observation (feature) vector $X$  taking values in $\R^d$. Given a measurable predictor $f:\R^d \to \Y$ and measurable  loss function $\ell: \Y\times \Y\to \R_+$, the loss incurred is  $\ell(Y,f(X))$. The minimum expected risk in predicting $Y$ from the random vector $X$  is
\[
 L_\ell^*(Y|X) = \inf_{f:\R^d\to \Y} \EXP[\ell(Y,f(X))],
\]
where the infimum is over all measurable $f$.

Suppose  that the tasks of collecting data and making the prediction are separated in time or in space. For example, the separation in time happens, when first the data are collected and  the statistical modelling and analysis are made much later. Separation in space can be due, for example, to collecting data at a remote location and making predictions centrally. Such situations are modeled by a transformation $T:\R^d \to \R^{d'}$ so that the prediction regarding $Y$ is made from the transformed observation $T(X)$, instead of $X$. An important example for such a transformation is quantization, in which case $T(X)$ is a discrete random variable.  Clearly, one always has $L_\ell^*(Y|T(X)) \ge  L_\ell^*(Y|X)$. The difference $L_\ell^*(Y|T(X)) - L_\ell^*(Y|X)$ is sometimes referred to in the literature as excess risk. A part of this paper is concerned with transformations for which the excess risk is zero no matter what the underlying loss function $\ell$ is. Such transformations are universally lossless in the sense that they can be chosen before the cost function $\ell$ for the underlying problem is known. More formally we make the following definition.

\begin{defin}[lossless transformation]
  \label{def1}
For a fixed joint distribution of $Y$ and $X$, a (measurable) transformation $T:\R^d\to
\R^{d'}$  is called \emph{universally lossless} if for any loss
function  $\ell: \Y\times \Y\to \R_+$ we have 
\[
L_\ell^*(Y|T(X)) = L_\ell^*(Y|X).
\]
\end{defin}

An important special transformation is feature selection.
Formally, for the observation (feature) vector  $X=(X^{(1)},\dots ,X^{(d)})$ and  $S\subset \{1,\dots ,d\}$, consider the $|S|$-dimensional vector $X_S=(X^{(i)}, i\in S)$. Typically, the dimension $|S|$ of $X_S$ is significantly smaller than $d$, the dimension of $X$. 
If we have
\[
L_\ell^*(Y|X_S) = L_\ell^*(Y|X),
\]
for all loss functions $\ell$, then the feature selector $X\mapsto X_S$ is  universally lossless. For fixed loss $\ell$, the performance of any statistical inference method is sensitive to the dimension of the feature vector. Therefore, dimension reduction is crucial before choosing or constructing such a method.
If $X_S$ universally lossless, then the complement feature subvector $X_{S^c}$
is irrelevant. It is an  open research problem  how to efficiently search a  universally lossless $X_S$ with minimum size~$|S|$. Since typically the distribution of the pair $X$ and $Y$ is not known and must be inferred from data, any such search algorithm  needs a procedure for testing for the  universal losslessness property of a feature selector. 

In the first part of this paper we give  a necessary and sufficient condition for a given transformation $T$ to be universally lossless and then  construct a partitioning based  statistic for testing this condition if independent and identically distributed training data are available. 
With the null hypothesis being that a given transformation is universally lossless, the test is shown to be strongly consistent in the sense that almost surely (a.s$.$) it makes finitely many Type I and II errors.

In many situations requiring that a transformation $T$ be universally lossless may be too demanding. The next definition relaxes this requirement.

\begin{defin}[$\delta$-lossless transformation]
  \label{def2}
For a fixed joint distribution of $Y$ and $X$, and $\delta>0$,  a transformation $T:\R^d\to
\R^{d'}$  is called \emph{universally $\delta$-lossless} with respect to a 
 class of loss functions $\L$, if we have 
\[
L_\ell^*(Y|T(X)) -L_\ell^*(Y|X) \le \delta
\quad \text{for all $\ell\in \mathcal{L}$.}
\] 
\end{defin}

In the second part of the paper we derive bounds on the excess minimum risk $L_\ell^*(Y|T(X)) -L_\ell^*(Y|X)$ in terms of the mutual information difference $I(Y;X)-I(Y;T(X))$ under various assumptions on $\ell$. 
With the aid of these bounds, we give information-theoretic sufficient conditions for a transformation $T$ to be $\delta$-lossless with respect to fairly general classes of loss functions~$\ell$. Applications to classification, nonparametric regression, portfolio strategies, the information bottleneck method,  and deep learning are also reviewed. 

\medskip 

\noindent\emph{Relationship with prior work:}  

Our first result, Theorem~\ref{thm2}, which shows that a
transformation is universally lossless if and only if it is a
sufficient statistic, is likely known, but we could not find it in
this explicit form in the literature (a closely related result is the
classical Rao-Blackwell theorem of mathematical statistics, e.g.,
Schervish \cite[Theorem~3.22]{Sch95}). Due to this result, testing
from independent data whether or not given a transformation is
universally lossless turns into a test for conditional
independence. Our test in Theorem~\ref{coro0} is based on the main
results in Gy\"orfi and Walk \cite{GyWa12}, but our construction is
more general and we also correct an error in the proof of
\cite[Theorem~1]{GyWa12}. Apart from \cite{GyWa12}, most of the
results in the literature for testing for conditional independence are
for real valued random variables and/or assume certain special
distribution types; typically the existence of a joint probability
density function. Such assumptions exclude problems where $Y$ is
discrete and $X$ is continuous, as is typical in classification, or
problems where the observation $X$ is concentrated on a lower
dimensional subspace or manifold. In contrast, our test construction
is completely distribution free and its convergence properties are
also (almost) distribution free. A more detailed review of related
work is given in Section~\ref{sec-univlossless}.

The main result in Section~\ref{sec-delta-lossless} is Theorem~\ref{thm-subg}, which  bounds the excess risk in terms of the square root of the mutual information difference $I(Y;X)-I(Y;T(X))$.  There is a history of such bounds starting possibly with Xu and Raginsky \cite{XuRa17},  where the generalization error of a learning algorithm was upper bounded by constant times the square root of the mutual information between the hypothesis and the training data (see also the references in \cite{XuRa17} and \cite{XuRa22}). This result has since been extended in various forms, mostly concentrating on providing information-theoretic bounds on the generalization capabilities of learning algorithms instead of looking at the excess risk; see, e.g., Raginsky et al.\ \cite{RaRaXu21}, Lugosi and Neu \cite{LuNe22}, Jose and Simeone \cite{JoSi21}, and the references therein, just to  mention a few of these works. The most relevant recent work relating to our bounds in Section~\ref{sec-delta-lossless} seems to be Xu and Raginsky  \cite{XuRa22},  where, among other things, information-theoretic bounds were developed on the  excess risk in a Bayesian learning framework; see also Hafez-Kolahi et al.\ \cite{HaMoKa23}.  The bounds in \cite{XuRa22} are not  on the excess risk $L_\ell^*(Y|T(X)) - L_\ell^*(Y|X)$; they involve training data, but their  forms are similar to ours. It appears that our Theorem~\ref{thm-subg} gives a  bound that holds uniformly for a larger class of loss functions $\ell$  and joint distributions of $Y$ and $X$; however, in \cite{XuRa22} several other bounds are presented that are tighter and/or allow more general distributions, for specific, fixed, loss functions.

\medskip 

\noindent\emph{Organization:}  

The paper is organized as follows. In Section~\ref{sec-lossless} we characterize universally lossless transformations and introduce a novel strongly consistent test for the universal losslessness property.
In Section~\ref{sec-delta-lossless} information-theoretic bounds on the excess minimum risk are developed and are used to characterize the $\delta$-losslessness property of transformations. Section~\ref{sec-appl} surveys connections with and applications to specific prediction problems as well as the information bottleneck method in deep learning. The somewhat lengthy  proof of the strong consistency of the test in Theorem~\ref{coro0} is given in Section~\ref{sec-proof}. 
Concluding remarks are given in Section~\ref{sec-concl}.

\section{Testing the universal losslessness property}

\label{sec-lossless}

In this section, we first give a characterization of universally lossless transformations for a given distribution of the pair $(X,Y)$. In practice, the distribution of $(X,Y)$ may not be known, but a sequence of independent and identically distributed (i.i.d$.$) copies of $(X,Y)$ may be available. For this case, we construct a procedure to test if a given transformation is universally  lossless and prove that, under mild conditions,  the test is strongly consistent. 

\subsection{Universally lossless transformations}

\label{sec-univlossless}

Based on Definition~\ref{def1}, introduce  the null hypothesis
\begin{align}
\label{hyp0}
\mathcal H_0=\{T:\mbox{ the transformation } T \mbox{ is universally lossless.}\}  
\end{align}

A transformation (statistic) $T(X)$ is called sufficient if the random variables  $Y$,  $T(X)$, $X$ form a Markov chain in this order, denoted by $Y\to T(X)\to X$ (see, e.g., Definition~3.8 and Theorem 3.9 in Polyanskiy and Wu \cite{PoWu22}). 

For binary valued $Y$, Theorems 32.5 and 32.6 in Devroye et al. \cite{DeGyLu96} imply that the statistic $T(X)$ is universally lossless if and only if it is sufficient. The following theorem extends this property to general $Y$. The result is likely known, but we could not find it in the given form.

\begin{theorem}
\label{thm2}
The transformation $T$ is universally
lossless if and only if  $Y\to T(X) \to X$ is a Markov chain.
\end{theorem}

\begin{proof} Assume first that  $Y\to T(X) \to X$ is a Markov chain.  This is
  equivalent to having   $\PROB(Y\in A| X,T(X)) = \PROB(Y\in A|T(X)) $ almost surely (a.s.) for any measurable $A\subset \Y$. Then we have
\begin{align*}
L_\ell^*(Y|X,T(X))
&=  \EXP\big[\inf_{y\in \Y} \EXP [\ell(Y,y)|X,T(X)] \big]  \\
&= \EXP\big[\inf_{y\in \Y} \EXP [\ell(Y,y)|T(X)] \big]  \\
&=  L_\ell^*(Y|T(X)).
\end{align*}
Since $L_\ell^*(Y|X,T(X))\le  L_\ell ^*(Y|X)\le L_\ell^*(Y|T(X))$ always holds,
we obtain  $L_\ell^*(Y|T(X))= L_\ell ^*(Y|X)$ for all $\ell$, so  $T(X)$ is universally
lossless.\\
Now assume that that the Markov chain condition $Y\to T(X)\to X$ does not
hold. Then there exist measurable $A\subset \Y$ with $0<\PROB(Y\in A)<1$ and
$B\subset \R^d$ with $\PROB(X\in B)>0$  such that
\[
  \PROB(Y\in A|X,T(X)) \neq \PROB(Y\in A| T(X)) \quad \text{if $X\in B$}.
\]
Let $h(y)=\IND_{y\in A}$, where $\IND_E$ is the indicator function of event $E$, and define the binary valued $\hat{Y}$ as
$\hat{Y}=h(Y)$. Then the Markov chain condition $\hat{Y}\to T(X)\to X$ does not
hold.  For this special case, Theorems~32.5 and 32.6 in
\cite{DeGyLu96} show that that there exist a loss function
$\hat{\ell}: \{0,1\}^2\to \R_+$ such that
$L_{\hat{\ell}}^*(\hat{Y}|T(X))> L_{\hat{\ell}} ^*(\hat{Y}|X)$. Finally, letting
$\ell(y,y') = \hat{\ell}(h(y),h(y'))$, we have
\[
  L_{\ell}^*(Y|T(X))=  L_{\hat{\ell}}^*(\hat{Y}|T(X))> L_{\hat{\ell}}^*(\hat{Y}|X)=  L_{\ell}^*(Y|X),
\]
which shows that $T(X)$ is not universally lossless.
\end{proof}

\subsection{A strongly consistent test}

Theorem~\ref{thm2} implies an equivalent form of the losslessness null hypothesis defined by (\ref{hyp0}):
\begin{align}
\label{hyp1}
\mathcal{H}_0=\{T:\, Y\to T(X) \to X \mbox{ is a Markov chain}\}, 
\end{align}
or equivalently, $\mathcal{H}_0$ holds if and only if $X$ and $Y$ are conditionally independent given $T(X)$:
\[
\mathcal{H}_0: \PROB\big(X\in A, Y\in B \mid T(X)\big)  =\PROB\big(X\in A \mid T(X)\big) \PROB\big(Y\in B \mid T(X)\big)  \quad \text{a.s.}
\]
for arbitrary Borel sets $A,B$. 
Furthermore, we consider the general case where the alternative hypothesis $\mathcal{H}_1$ is the complement of $\mathcal{H}_0$:  $\mathcal H_1=\mathcal H_0^c$.

Now assume that the joint distribution of $(X,Y,T(X))$ is not known, but instead a 
 sample of independent and identically distributed (i.i.d$.$) random vectors
$(X_1,Y_1,Z_{1}),\dots ,(X_n,Y_n,Z_{n})$ having  common distribution that of  $(X,Y,Z)$ is given, where $Z_i=T(X_i)$ and $Z=T(X)$. The goal is to test the hypothesis $\mathcal{H}_0$ of conditional independence based on this data.

For testing conditional independence, most of the results in the literature are on real valued $X,Y,Z$.
Based on kernel density estimation, Cai et al.\ \cite{CaLiZh22} introduced a test statistic and 
under the null hypothesis  calculated its limit distribution.
In Neykov et al.\ \cite{NeBaWa21} a gap is introduced between the null and alternative hypotheses. This gap is characterized by the total variation distance, which decreases with increasing $n$. Under some smoothness conditions, minimax bounds were derived.
According to Shah and  Peters \cite{ShPe20}, a regularity condition such as our  Lipschitz condition \eqref{Lip} below cannot be omitted  if a test for conditional independence is to be consistent.
This is a consequence of their No-Free Lunch Theorem that states
under general conditions that if under the null hypothesis the bound
on the error probability is non-asymptotic, then under the alternative
hypothesis the rate of convergence of the error probability can be
arbitrarily slow, which is a well known phenomenon in nonparametric
statistics. We note that these cited results,  and indeed most of the results in the literature for testing for conditional independence,  are for real valued random variables and/or assume certain special distribution types , typically  the existence of a joint probability density function or that both $X$ and $Y$ are discrete, as in \cite{NeBaWa21}. As we remarked earlier, such  assumptions exclude problems where  $Y$ is discrete and $X$ is continuous (typical in classification), or problems  where the observation $X$ is concentrated on a lower dimensional subspace or manifold. In contrast, our test construction is completely distribution free and its convergence properties are almost distribution free (we do  assume a mild Lipschitz-type condition; see the upcoming  Condition~\ref{cond1})

In our hypotheses testing setup, the alternative hypothesis, $\mathcal{H}_1$, is the complement of the null hypothesis,  $\mathcal{H}_0$; 
therefore there is no separation gap between the hypotheses.
Dembo and Peres \cite{DePe94} and Nobel \cite{Nob06} characterized
hypotheses pairs that admit strongly  consistent tests, i.e.,  tests
that, with probability one, only make finitely many Type I and II
errors. 
This property is called discernibility.
As an illustration of the intricate nature of the discernibility concept,
Dembo and Peres \cite{DePe94} demonstrated an exotic example, where the null hypothesis is that the mean of a random variable is rational, while the alternative hypothesis is that this mean minus $\sqrt{2}$ is rational.
(See also  Cover \cite{Cov73} and Kulkarni and Zeitouni \cite{KuZe91}.)
The discernibility property shows up in
Biau and Gy\"orfi \cite{BiGy05} (testing homogeneity),
Devroye and Lugosi \cite{DeLu02} (classification of densities),
Gretton and Gy\"orfi \cite{GrGy10} (testing independence),
Morvai and Weiss \cite{MoWe21} and Nobel \cite{Nob06} (classification of stationary processes),
among others. 

In the rest of this section, under mild conditions on the distribution of $(X,Y)$, we study discernibility in the context of lossless transformations for statistical inference with general risk. We will  make strong use of the multivariate partitioning based test of  Gy\"orfi and Walk \cite{GyWa12}.

Let $P_{XYZ}$ denote  the joint distribution of $(X,Y,Z)$  and similarly for any marginal distribution of  $(X,Y,Z)$; e.g., $P_{XZ}$ denotes the distribution of the pair $(X,Z)$. 
As in Gy\"orfi and Walk \cite{GyWa12}, introduce the following empirical distributions:
\begin{spreadlines}{2ex} 
\begin{align*}
P^n_{XYZ}(A,B,C) &=\frac{\#\{i : (X_i,Y_i,Z_i) \in A\times B\times C, i=1,
\ldots,n\}}{n}, \\
P^n_{XZ}(A,C)& =\frac{\#\{i : (X_i,Z_i) \in A\times C, i=1,
\ldots,n\}}{n}, \\ 
P^n_{YZ}(B,C) & = \frac{\#\{i : (Y_i,Z_{i}) \in B\times C, i=1,
\ldots,n\}}{n}, \\
\intertext{and} 
P^n_{Z}(C)& =\frac{\#\{i : Z_{i} \in C, i=1, \ldots,n\}}{n},
\end{align*} 
\end{spreadlines}
for Borel sets $A\subset \R^d$, $B\subset \R$ and $C\subset \R^{d'}$.

For the sake of simplicity, assume that $X$, $Y$ and $Z=T(X)$ are bounded.
Otherwise, we apply a componentwise, one-to-one scaling into the interval $[0,1]$. 
Obviously, the losslessness null hypothesis $\mathcal H_0$ is invariant under such scaling.
Let
\[
\P _n=\{A_{n,1},\ldots ,A_{n,m_n}\}, \; \Q _n=\{B_{n,1},\ldots ,B_{n,m'_n}\}, \; {\cal R}_n=\{C_{n,1},\ldots ,C_{n,m''_n}\}
\]
be  finite cubic partitions of the ranges of $X$, $Y$ and $Z$  with all the cubes having common side lengths  $h_n$ (thus $h_n$ is proportional to $1/m_n'$). As in  \cite{GyWa12} we define the test statistic  
\begin{align}
\label{Ln}
L_n
&= \sum_{A\in \P_n,B\in \Q_n,C\in {\cal R}_n}
\left|P^n_{XYZ}(A,B,C)
-\frac{P^n_{XZ}(A,C)P^n_{YZ}(B,C)}{P^n_{Z}(C)}
\right|.
\end{align}
Our test rejects $\mathcal{H}_0$  if 
\[
L_n \ge t_n, 
\]
and accepts it if $L_n< t_n$, where the threshold $t_n$ is set to
\begin{align}
\label{eq-tndef}
t_n &= 
c_1\left(\sqrt{\frac{m_nm'_nm''_n}{n}} + \sqrt{\frac{m'_nm''_n}{n}}
+ \sqrt{\frac{m_nm''_n}{n}}+\sqrt{\frac{m''_n}{n}}\right)+(\log n)h_n, 
\end{align}
where the constant $c_1$ satisfies 
\[
c_1>\sqrt{2\log 2}\approx 1.177.
\]

In this setup the distribution of $(X,Y)$ is arbitrary; its components can be discrete or absolutely continuous or the mixture of the two or even singularly continuous.
It is important to note that for constructing this test, there is no need to know the type of the distribution.

We assume that the joint distribution of $X$, $Y$, and $Z=T(X)$  satisfies the following assumption.

\begin{condition}
\label{cond1}
Let $p(\,\cdot \,| z)$ be the density of the conditional distribution $P_{X|Z=z}=\PROB(X\in \cdot\;|Z=z)$ with respect to the distribution  $P_X$ as a dominating measure and introduce the notation
\[
C_n(z)=C_{n,j} \mbox{ if } z\in C_{n,j}.
\]
Assume  that for some $C^*>0$, $p(x| z)$ satisfies the  condition
\begin{align}
\label{Lip}
\int\int   \bigg|  p(x| z)- \frac{\int_{C_n(z) }p(x|z')P_Z(dz')}{P_Z(C_n(z))}\bigg| \, P_X(dx) P_Z(dz)
&\le
C^*  h_n,
\end{align}
for all $n$.
\end{condition}

We note that the ordinary Lipschitz condition
\begin{align}
\label{TVLip}
\int \big| p(x|z)-p(x| z')\big|\, P_X(dx)
&\le
\frac{C^*}{\sqrt{d'}}\|z-z'\| \quad \text{for all $z,z'\in \R^{d'}$}
\end{align}
implies (\ref{Lip}). This latter condition is equivalent to 
\[
d_{TV}\big(P_{X|Z=z},P_{X|Z=z'}\big) \le \frac{C^*}{2\sqrt{d'}}\|z-z'\| \quad \text{for all $z,z'\in \R^{d'}$}, 
\]
where  $d_{TV}(P,Q)$ denotes the total variation distance between distributions $P$ and $Q$. 
In Neykov,   Balakrishnan and  Wasserman \cite{NeBaWa21}, condition  (\ref{TVLip})
is called the Null TV Lipschitz condition.

The next theorem is an adaptation and extension of the results in Gy\"orfi and Walk \cite{GyWa12} to this particular problem of  lossless transformation. In \cite{GyWa12} it was assumed that the  sequence of partitions $\{\P _n,\Q _n,{\cal R}_n\}$ is nested, while we make no such assumption.  The proof, in which an error made in \cite{GyWa12} is also corrected,  is relegated to Section~\ref{sec-proof}. 

\begin{theorem}
\label{coro0}
Suppose that $X$, $Y$ and $Z=T(X)$ are bounded and Condition~\ref{cond1} holds
for all $n$. If the   sequence $h_n$ satisfies 
\begin{equation}
\label{hnS1}
\lim _{n\to \infty}nh_n^{d+1+d'}= \infty
\end{equation}
and
\begin{align}
\label{hnS2}
\lim_{n\to \infty}h_n^{d'}\log n = 0,
\end{align}
then we have following: 
\begin{itemize}
\item[\rm{(a)}] 
Under the losslessness null hypothesis ${\mathcal H}_0$,
we have for all  $n\ge e^{C^*}$,
\begin{equation} 
\label{eq-summable}
\PROB(L_n \ge t_n) \le 4e^{-(c_1^2/2-\log 2) m''_n},
\end{equation}
and therefore, because $\sum_{n=1}^{\infty} \PROB(L_n \ge t_n)  <\infty$ by \eqref{hnS2} and \eqref{eq-summable},  after a random sample size,  the test makes  no error with probability one. 
\item[\rm{(b)}] 
Under the alternative hypothesis ${\mathcal H}_1={\mathcal H}_0^c$,
\[
\liminf_{n\to \infty} L_n>0  \quad {\text a.s.,}
\]
and so, with probability one,  after a random sample size the test makes no error. 
\end{itemize}
\end{theorem}

\noindent \emph{Remarks:} 
\vspace{-3pt}

\begin{enumerate}[label=(\roman*)] 
\item  
The  choice $h_n=n^{-\delta}$ with  $0<\delta< 1/(d+1+d')$  satisfies both conditions \eqref{hnS1} and~$\eqref{hnS2}$. 

\item Note that by \eqref{eq-tndef},  $t_n$ is of order $c_1 \sqrt{\frac{m_nm'_nm''_n}{n}}+(\log n)h_n$.
Since we have 
\begin{align*}
m_n=O(1/h_n^d), \quad m'_n=O(1/h_n), \quad m''_n=O(1/h_n^{d'}),
\end{align*}
this means that  $t_n$  is of order $\sqrt{1/(nh_n^{d+1+d'})}+(\log n)h_n$.

\end{enumerate}

An important special transformation is given by the feature selection $X_S$ defined in the Introduction.
Theorem \ref{coro0} demonstrates the possibility of universally lossless dimension reduction for any multivariate feature vector.
Note that in the setup of feature selection, the partition ${\cal P}_n$ can be the nested version of ${\cal R}_n$ and so the calculation of the test statistic $L_n$ is easier.

\section{Universally $\delta$-lossless transformations}

Here we develop  bounds on the excess minimum risk in terms of  mutual information under various assumptions on the loss function. 
With the aid of these bounds, we give information-theoretic sufficient conditions for a transformation $T$ to be universally $\delta$-lossless with respect to fairly general classes of loss functions~$\ell$.

\label{sec-delta-lossless}

\subsection{Preliminaries on mutual information}

\label{sec-pre}

Let $P_{XY}$ denote the joint distribution of the pair $(X,Y)$ and let $P_XP_Y$
denote the product of the marginal distributions of $X$ and $Y$, respectively. 
The mutual information between $X$ and $Y$, denoted by  $I(X;Y)$, is
defined as
\[
   I(X;Y) = D(P_{XY}\|P_XP_Y),
\]
where 
\[
D(P\|Q) = \begin{cases}
    \int \frac{dP}{dQ} \log \big( \frac{dP}{dQ} \big) \, dQ & \text{ if $P\ll Q$} \\
       \infty & \text{ otherwise,} 
\end{cases}
\]
is the Kullback-Leibler (KL) divergence between probability
distributions $P$ and $Q$ (here $P\ll Q$ means that $P$ is absolutely continuous with respect to $Q$ with Radon–Nikodym derivative $\frac{dP}{dQ}$). Thus $I(X;Y)$ is always nonnegative  and $ I(X;Y) =0$
if and only if $X$ and $Y$ are independent (note that  $
I(X;Y)=\infty$ is possible). In this  definition  and throughout the paper $\log$ denotes the natural logarithm.

For random variables $U$ and $V$ (both taking values in finite-dimensional
Euclidean spaces), let $P_{U|V}$ denote the conditional
distribution of $U$ given $V$. Furthermore, let $P_{U|V=v}$ denote the stochastic kernel
(regular conditional probability) induced by $P_{U|V}$. Thus, in particular,
$P_{U|V=v}(A) =\PROB(U\in A|V=v)$ for each measurable set $A$.

Given another random variable $Z$, the conditional mutual information $I(X;Y|Z)$ is defined as
\[
I(X;Y|Z)  = \int D(P_{XY|Z=z}\|P_{X|Z=z}P_{Y|Z=z}) P_Z(dz).
\]
The  integral above is also denoted by $D(P_{YX|Z}\|P_{Y|Z} P_{X|Z}| P_Z)$ and
is called the conditional KL divergence. One can define
\[
I(X;Y|Z=z)=  D(P_{XY|Z=z}\|P_{X|Z=z}P_{Y|Z=z}) 
\]
so that
\begin{equation}
  \label{cond-inf-def}
I(X;Y|Z)  = \int I(X;Y|Z=z) P_Z(dz).
\end{equation}
From the definition it is clear that
$I(X;Y|Z)=0$ if and only if $X$ and $Y$ are conditionally independent given $Z$,
i.e., if and only if $Y\to Z\to X$ (or equivalently, if and only if  $X\to Z\to Y$).

Another way of expressing $I(X;Y)$ is
\begin{equation}
\label{idefalt}
  I(X;Y)= \int D(P_{Y|X=x}\| P_{Y}) P_X(dx). 
\end{equation}
One can check that in a similar way  $I(X;Y|Z)$ can be expressed
as
\begin{equation}
  \label{cond-inf-def1}
 I(X;Y|Z)  =     \iint  D(P_{Y|X=x,Z=z}\| P_{Y|Z=z}) P_{X|Z=z}(dx) P_Z(dz) .
\end{equation}

Properties of mutual information and conditional
mutual information, their connections to the KL divergence, and identities
involving these information measures are detailed in, e.g., 
Cover and Thomas \cite[Chapter~2]{CoTh06} and
Polyanskiy and Wu \cite[Chapter~3]{PoWu22}.

\subsection{Mutual information bounds and  $\delta$-lossless transformations}


A real random variable $U$ with finite expectation is said to be $\sigma^2$-subgaussian for some $\sigma^2> 0$ if 
\[
\log \EXP\big[e^{\lambda (U-\EXP[U]) }\big] 
    \le  \dfrac{\sigma^2 \lambda^2}{2} \quad \text{for all $\lambda\in \R.$}
\]
Furthermore, we say that $U$ is conditionally $\sigma^2$-subgaussian given another random variable $V$ if we have a.s. 
\begin{equation}
\label{eq-cond-subg}  
\log \EXP\big[e^{\lambda (U-\EXP[U\,|\, V] )} \,\big|\,V\big] 
    \le     \dfrac{\sigma^2 \lambda^2}{2}  \quad \text{for all $\lambda\in \R$.}
\end{equation}

The following result gives a quantitative upper bound on the excess minimum risk
$ L_{\ell}^*(Y|T(X))- L_{\ell}^*(Y|X)$ in terms of  the mutual information difference $ I(Y;X)-I(Y;T(X))$ under certain,
not too restrictive, conditions.  Note that $ L_{\ell}^*(Y|T(X))- L_{\ell}^*(Y|X)\ge 0$ always holds.

Given   $\epsilon>0$, we call an estimator
$f': \R^d\to \Y$ $\epsilon$-optimal if  $\EXP[\ell(Y,f'(X))] <
L_\ell^*(Y|X)+\epsilon$.

\begin{theorem}
  \label{thm-subg}
 Let $T:\R^d\to \R^{d'}$ be a measurable transformation and assume
 that for any $\epsilon>0$, there exists an $\epsilon$-optimal
 estimator $f'$ of $Y$ from $X$  such that   $\ell(y,f'(X)))$ is
 conditionally $\sigma^2(y)$-subgaussian  given $T(X)$ for every $y\in \Y$, i.e.,
\begin{equation} 
\label{eq-subgy}
 \log \EXP\big[e^{\lambda \big(\ell(y,f'(X))-\EXP[\ell(y,f'(X)) \,\big|\,   T(X)] \big)} \mid T(X)\big] 
    \le     \dfrac{\sigma^2(y) \lambda^2}{2}  \quad \text{a.s}
\end{equation} 
for all $\lambda\in \R$ and $y\in \R$,   where   $\sigma^2: \R\to \R_+$  satisfies 
$\EXP[\sigma^2(Y)]<\infty$. Then, one has 
\begin{align}  
L_\ell^*(Y|T(X))- L_\ell^*(Y|X)   
&\le    \sqrt{ 2\EXP[\sigma^2(Y)] I(Y;X\,|\, T(X))  } \nonumber \\
& = \sqrt{2\EXP[\sigma^2(Y)]\big( I(Y;X)-I(Y;T(X)) \big) }\; .  \label{eq2}  
 \end{align}
\end{theorem}

\noindent \emph{Remarks:} 
\vspace{-3pt}
\begin{enumerate}[label=(\roman*)]
\item
In case $I(Y;X|T(X)) =\infty$, we interpret the right hand side of \eqref{eq2} as~$\infty$. With this interpretation, the bound always holds. 

\item  We show in Section~\ref{sec-regr} that  the subgaussian
  condition \eqref{eq-subgy} holds for the regression problem with
  squared error $\ell(y,y')=(y-y')^2$ if  $Y=m(X)+N$, where  $N$   is
  independent noise having zero mean and finite fourth moment
  $\EXP[N^4]<\infty$, and the regression function $m(x)=E[Y|X=x]$ is
  bounded. In particular, the bound in the theorem holds if $N$ is
  normal with zero mean and $m$ is bounded.

\item Although hidden in the notation, $\EXP[\sigma^2(Y)]$ depends on the loss function $\ell$. Thus the upper bound \eqref{eq2}   is the product of two terms, the second of which, $\sqrt{ I(Y;X)-I(Y;T(X)) }$, is independent of the loss function. 

\item The bound in the theorem is not tight in general. In Section~\ref{sec-portf} an example is given in the context of portfolio selection, where the excess risk can be upper bounded by the difference $I(Y;X)-I(Y;T(X))$.

\item
The proof of Theorem~\ref{thm-subg} and those of its
corollaries go  through virtually without
change if we replace $T(X)$ with  any $\R^{d'}$-valued random variable $Z$  such that
$Y\to X\to Z$.   Under the conditions of the theorem, we then have 
\begin{align*}  
 L_\ell^*(Y|Z) - L_\ell^*(Y|X)   
&\le    \sqrt{ 2\EXP[\sigma^2(Y)]I(Y;X|Z)  } \nonumber \\
& = \sqrt{2\EXP[\sigma^2(Y)]\big( I(Y;X)-I(Y;Z) \big)}\; .
\end{align*}

In fact,  Theorem~\ref{thm-subg} and its
corollaries  hold for general random variables $Y$, $X$, and $Z$ taking values in complete and separable metric (Polish) spaces $\mathcal{Y}$,  $\mathcal{X}$, and   $\mathcal{Z}$, respectively, if $Y\to X\to Z$.
\end{enumerate}

The proof of Theorem~\ref{thm-subg} is based on a slight generalization of  Raginsky et al.\ \cite[Lemma~10.2]{RaRaXu21},  which we state next. In the lemma,  $U$ and $V$ are arbitrary abstract random variables defined on the same probability space and taking values in spaces  $\U$ and $\V$, respectively,  $\bar{U}$ and $\bar{V}$ are independent copies of $U$ and $V$ (so that $P_{\bar{U}\bar{V}}= P_UP_V$), and $h:\U\times \V\to \R$ is a measurable function.

\begin{lemma}
\label{lem-RagRakXulemma}
Assume that $h(u,V)$ is $\sigma^2(u)$-subgaussian for all $u\in \U$, where $\EXP[\sigma^2(U)]<\infty$. Then 
\[
\big| \EXP[h(U,V)] - \EXP[h(\bar{U},\bar{V})] \big| \le 
\sqrt{2 \EXP[\sigma^2(U)] I(U;V)}\; .
\]
\end{lemma}

\begin{proof}
We essentially copy the proof of    \cite[Lemma~10.2]{RaRaXu21} where it was assumed that $\sigma^2(u)$ does not depend on $u$. With this restriction, the subgaussian condition \eqref{eq-subgy} in Theorem~\ref{thm-subg} would have to hold with $\sigma^2(y)\le \sigma^2$ uniformly over $y$. This condition  would exclude regression models with independent subgaussian noise and, \emph{a fortiori},  models with independent noise that does not possess finite absolute moments of all orders , while our Theorem~2 can also be applied  in such cases  (see Section~\ref{sec-regr})

We make use of the Donsker–Varadhan variational representation of the relative entropy \cite[Corollary~4.15]{BuLuMa13}, which states that 
\[
D(P\|Q) = \sup_F \left(  \, \int F dP - \log \int e^F dQ \, \right),
\]
where the supremum is over all measurable $F:\Omega\to \R$ such that $\int e^FdQ    <\infty$. Applying this with $P=P_{V|U=u}$, $Q=P_V$, and $F=\lambda h(u,V)$, we obtain
\begin{align}
D(P_{V|U=u}\|P_V) 
&\ge 
\EXP[\lambda h(u,V)|U=u]- \log \EXP[e^{\lambda h(u,V)}]      \nonumber \\
& \ge  
\lambda\big( \EXP[ h(u,V)|U=u]- \EXP[\lambda h(u,V)] \big) -\frac{\lambda^2\sigma^2(u)}{2}, \label{eq-dbound}
\end{align}
where the second inequality follows from assumption that  $h(u,V)$ is $\sigma^2(u)$-subgaussian.  Maximizing the right hand side of  \eqref{eq-dbound} over  $\lambda\in \R$ gives, after rearrangement, 
\begin{equation}
 \label{eq-dupper}   
 \big| \EXP[ h(u,V)|U=u]- \EXP[ h(u,V)] | \le 
 \sqrt{2 \sigma^2(u)  D(P_{V|U=u}\|P_V)}. 
\end{equation}
Since $\bar{U}$ and $\bar{V}$ are independent,  $ \EXP[ h(u,V)]= \EXP[h(\bar{U},\bar{V})|\bar{U}=u]$, and we obtain 
\begin{align}
\MoveEqLeft \big| \EXP[h(U,V)] - \EXP[h(\bar{U},\bar{V})] \big|\nonumber  \\
&=
\bigg| \int \big( \EXP[h(U,V)|U=u] -\EXP[ h(\bar{U},\bar{V})|\bar{U}=u]\big) \,P_U(du) \bigg| \nonumber \\
&= \bigg| \int \Big( \EXP[h(u,V)|U=u] - h(u,V) \Big) \,P_U(du) \bigg| \nonumber \\
&\le \int \big| \EXP[h(u,V)|U=u] - h(u,V) \big|  \,P_U(du)   \label{eq-jen}\\
&\le  \int \sqrt{2 \sigma^2(u)  D(P_{V|U=u}\|P_V)}  \,P_U(du)  \label{eq-dupper1}   \\
&\le  \sqrt{ \int 2 \sigma^2(u) \,P_U(du)  } \; \sqrt{ \int  D(P_{V|U=u}\|P_V)  \,P_U(du)}    \label{eq-cauchy} \\
&= \sqrt{2 \EXP[\sigma^2(U)] I(U;V)},  
\end{align}
where \eqref{eq-jen} follows from Jensen's inequality, \eqref{eq-dupper1}   follows from \eqref{eq-dupper},  in \eqref{eq-cauchy} we used the Cauchy-Schwarz inequality, and the last equality follows from \eqref{idefalt}.

\end{proof}

\begin{proof}[Proof of Theorem~\ref{thm-subg}]  
Let $\bar{Y}$ and $\bar{X}$  be random variables such that $P_{\bar{Y}|T(\bar{X})}=P_{Y|T(X)}$,  $P_{\bar{X}|T(\bar{X})} =P_{X|T(X)}$,  $P_{T(\bar{X})}= P_{T(X)}$, and $\bar{Y}$ and $\bar{X}$ are conditionally independent given $T(\bar{X})$. 
Thus the joint distribution of the triple $(\bar{Y},\bar{X},T(\bar{X}))$ is  $P_{\bar{Y}\bar{X}T(\bar{X})}= P_{Y|T(X)}P_{X|T(X)}P_{T(X)}$.

 We apply Lemma~\ref{lem-RagRakXulemma} with $U=Y$, $V=X$, and
 $h(u,v)=\ell(y,f'(x))$. Note that by the conditions of the theorem we
 can choose an  $\epsilon$-optimal $f'$ such for every  $y$,
 $\ell(y,f'(X))$ is conditionally  $\sigma^2(y)$-subgaussian given $T(X)$. Consider $\EXP [\ell(Y,f'(X))\,|\, T(X)=z]$ and $\EXP[\ell(\bar{Y},f'(\bar{X})) ,|\, T(\bar{X})=z]$ as regular (unconditional) expectations taken with respect to $P_{YX|T(X)=z}$ and 
$P_{\bar{Y}\bar{X}|T(\bar{X})=z}$ respectively,    and consider $I(Y;X|T(X)=z)$  as regular mutual information between random variables with distribution $P_{YX|T(X)=z}$.  Since   $\bar{Y}$ and $\bar{X}$ are conditionally independent  given
$T(\bar{X})=z$,  Lemma~\ref{lem-RagRakXulemma}  yields 
\begin{align*}
\MoveEqLeft[6]  \big| \EXP [\ell(Y,f'(X))\,|\, T(X)=z]  - \EXP [\ell(\bar{Y},f'(\bar{Y})) \,|\, T(\bar{X})=z] \big|\nonumber \\  
&\le \sqrt{2\EXP[\sigma^2(Y)|T(X)=z] I(X;Y|T(X)=z)}\;.
\end{align*}
Recalling that $T(\bar{X})$ and $T(X)$ have the same distribution and applying Jensen's inequality and the Cauchy-Schwarz inequality as in \eqref{eq-jen} and \eqref{eq-cauchy}, we obtain 
\begin{align}  
 \MoveEqLeft[0.5] \big| \EXP [\ell(Y,f'(X))]  - \EXP [\ell(\bar{Y},f'(\bar{X})) ] \big| 
\nonumber \\
& \le \int \big| \EXP [\ell(Y,f'(X))\,|\, T(X)=z]  - \EXP [\ell(\bar{Y},f'(\bar{X})) \,|\, T(\bar{X})=z] \big| \, P_{T(X)}(dz) \nonumber \\
&\le \int \sqrt{2\EXP[\sigma^2(Y)|T(X)=z] I(Y;X|T(X)=z)}  \, P_{T(X)}(dz) 
\nonumber \\
&\le 
 \sqrt{2\int \EXP[\sigma^2(Y)|T(X)=z]  \,P_{T(X)}(dz) } \; \sqrt{ \int I(Y;X|T(X)=z) \,P_{T(X)}(dz)}   \nonumber  \\
 & = \sqrt{2\EXP[\sigma^2(Y)]  I(Y;X|T(X))}\;. \label{eq3}  
\end{align}

\noindent  On the one hand, we have 
\begin{equation}
\label{eq_Tbound}
   \EXP \big[ \ell(\bar{Y};f'(\bar{X}))\big]  \ge L_\ell^*(\bar{Y}|\bar{X})  
   = L_\ell^*(\bar{Y}|T(\bar{X}))   
   =  L_\ell^*(Y|T(X)), 
\end{equation}
where the first equality follows from Theorem~\ref{thm2} by the conditional independence of  $\bar{Y}$ and $\bar{X}$ given $T(\bar{X})$,   and the second follows since  $(\bar{Y},T(\bar{X})) $ and $(Y,T(X))$ have the same distribution by construction.  
 On the other hand, $L_\ell ^*(Y|X) \ge \EXP[\ell(Y,f'(X))] - \epsilon $. Thus 
\eqref{eq3} and \eqref{eq_Tbound} imply  
\begin{align*}
0\le   L_\ell^*(Y|T(X))-L_\ell^*(Y|X) 
&\le  \EXP [\ell(\bar{Y},f'(\bar{X}))] - \EXP [\ell(Y,f' (X))]  +\epsilon \\
&\le   \sqrt{2\EXP[\sigma^2(Y)]  I(Y;X|T(X))} +\epsilon,
\end{align*}
which  proves the  upper bound in \eqref{eq2}  since  $\epsilon>0 $ is
arbitrary. By expanding $I(Y;X|Z)$ in two different ways using the
chain rule for mutual information (e.g., Cover and Thomas \cite[Thm.~2.5.2]{CoTh06}), and using the conditional independence of
$Y$ and $T(X)$ given $X$, one obtains $I(Y;X|T(X))= I(Y;X)-I(Y;T(X))$,
which shows the equality in \eqref{eq2}.
\end{proof}

\vspace{1em}

We state two corollaries for 
special cases. In the first, we assume that  $\ell$ is
  uniformly bounded, i.e.,  $\|\ell\|_{\infty} = \sup_{y,y'\in \Y}
  \ell(y,y')<\infty$. For any $c>0$, let $\L(c)$ denote the collection of all loss functions $\ell$ with $\|\ell\|_{\infty} \le c$. Recall the notion  of a universally $\delta$-lossless transformation from Definition~\ref{def2}.

\begin{cor}
  \label{cor1} Suppose the loss function $\ell$ is bounded. 
 Then  for any measurable  $T: \R^d\to \R^{d'}$,  we have  
\begin{eqnarray}
 L_\ell^*(Y|T(X))- L_\ell^*(Y|X)  & \le  \frac{\|\ell\|_{\infty}}{\sqrt{2}} \sqrt{
  I(Y;X)-I(Y;T(X))}\; .       \label{eq1}        
\end{eqnarray}
Therefore, whenever 
\begin{equation}
 \label{eq-ulosless1}   
  I(Y;X)-I(Y;T(X)) \le \frac{2\delta^2}{c^2},
\end{equation}
the transformation  $T$ is universally $\delta$-lossless  for the
family $\L(c)$,  i.e., $ L_\ell^*(Y|T(X))- L_\ell^*(Y|X) \le \delta $ for all $\ell$ with  $\|\ell\|_{\infty} \le c$.

\end{cor}

\noindent \emph{Remarks:} 
\vspace{-3pt}

\begin{enumerate}[label=(\roman*)] 
\item The bound of the theorem  can be used to
give an estimation-theoretic motivation of the information bottleneck (IB)
problem; see Section~\ref{sec-ib}.   
\item   Let $L_\ell^*(Y)= L_\ell^*(Y|\emptyset)= \inf_{y\in \Y}
\EXP[\ell(Y,y)] $. For bounded $\ell$, the inequality
\[
 L_\ell^*(Y)- L_\ell^*(Y|X)   \le 2\sqrt{2} \|\ell\|_{\infty}  \sqrt{I(Y;X) } 
\]
was proved in  Makhdoumi et \emph{al$.$} \cite[Theorem~1]{Mak-etal14}  for discrete
alphabets to motivate the so-called privacy funnel problem. This
inequality follows from \eqref{eq2} by setting $Z=T(X)$ to be constant there. 

\item A simple self-contained proof of   \eqref{eq1}  (see below) was provided by Or Ordentlich and communicated to the second author by Shlomo Shamai \cite{OrSh20}, in response to an early version of this manuscript.  
The bound in \eqref{eq1} seems to have first appeared in published form in  Hafez-Kolahi et al.\ \cite[Lemma~1]{HaMoKaBa21}, where the proof was attributed to Xu and Raginsky \cite{XuRa20}

\end{enumerate}

\begin{proof}[Proof of Corollary~\ref{cor1}]
 If $\ell$ is
  uniformly bounded, then for any   $f:\R^d\to \Y$ one has   $\ell(y,f(x)) \in [0,
  \|\ell\|_{\infty}]$ for all $y$ and $x$. Then 
   Hoeffding's lemma (e.g., Boucheron et al.\  
  \cite[Lemma~2.2]{BuLuMa13}) implies that for all $y$, 
  $\ell(y,f(X))$ is conditionally  $\sigma^2$-subgaussian with
  $\sigma^2= \frac{\|\ell\|_{\infty}^2}{4}$ given $T(X)$.  Since  an
  $\epsilon$-optimal estimator $f'$ exists for any $\epsilon>0$ and
  $\ell(y,f'(X))$ is conditionally 
  $\sigma^2$-subgaussian given $T(X)$ by the preceding argument, \eqref{eq1}   
  follows from Theorem~\ref{thm-subg}. The second statement follows directly from   \eqref{eq1}  and the fact that $\|\ell\|_{\infty}\le c$ for all $\ell\in \L(c)$. 

The following alternative argument by Or Ordentlich  \cite{OrSh20} is based on Pinsker's inequality on the total variation distance  in terms of the KL divergence (see, e.g.,  \cite[Theorem~7.9]{PoWu22}).  For bounded $\ell$, it gives a direct proof of an analogue of the key inequality \eqref{eq3} in the proof of Theorem~\ref{thm-subg}. This argument avoids Lemma~\ref{lem-RagRakXulemma}
 and the machinery introduced by the subgaussian assumption.
 
Using the same notation as in the proof of Theorem~\ref{thm-subg} and letting $P=P_{YXZ}$ and
  $Q=P_{\bar{Y}\bar{X}T(\bar{Y})}$,  we have 
  \begin{align*}
    \EXP \big[\ell(Y,f'(X))\big]  - \EXP \big[ \ell(\bar{Y},f'(\bar{X}))\big]  \big| &=                \int \ell(y,f'(x)) \, dP - \int \ell(y,f'(x)) \, dQ \nonumber\\
   & \le \|\ell\|_{\infty} d_{\rm TV}(P,Q) \nonumber  \\
 &\le   \frac{\|\ell\|_{\infty}}{\sqrt{2}}   \sqrt{D(P\|Q)
        } \qquad  \text{(by Pinsker's inequality)}  \nonumber  \\
      &= \frac{\|\ell\|_{\infty}}{\sqrt{2}}  \sqrt{D(P_{YX|T(X)}P_Z\|P_{Y|T(X)} P_{X|T(X)} P_{T(X)})} \nonumber \\   
    &= \frac{\|\ell\|_{\infty}}{\sqrt{2}}  \sqrt{D(P_{YX|T(X)}\|P_{Y|T(X)} P_{X|T(X)}| P_{T(X)})}  \nonumber \\
    &=  \frac{\|\ell\|_{\infty}}{\sqrt{2}}  \sqrt{I(X,Y|T(X))}\, . \nonumber 
  \end{align*}
 The rest of the proof proceeds exactly as in Theorem~\ref{thm-subg}. 
\end{proof}

In the second corollary we do not require that $\ell$ be bounded, but
assume that an optimal estimator $f^*_\ell$ from $X$ to $Y$ exists such
that $\ell(y,f^*_\ell(X))$ is conditionally  $\sigma^2(y)$-subgaussian given $T(X)$, where $\EXP[\sigma^2(Y)]<\infty$.

\begin{cor}
  \label{cor2}
  Assume that an optimal estimator $f^*_\ell$ of $Y$ from $X$ exist, i.e., the
  measurable function $f^*_\ell$ satisfies $\EXP[\ell(Y,f^*_\ell(X))] =
  L_\ell^*(Y|X)$. Suppose furthermore that the subgaussian condition of Theorem~\ref{thm-subg} holds with $f'=f^*_\ell$ (i.e., \eqref{eq-subgy}  holds for $f'=f^*_\ell)$. Then 
  \begin{align}
 L_\ell^*(Y|T(X))- L_\ell^*(Y|X)  &  \le \sqrt{2\EXP[\sigma^2(Y)] \big( I(Y;X)-I(Y;T(X)) \big) }\; .  \label{eq4}   
  \end{align}
\end{cor}

\begin{proof}
  The corollary immediately follows from Theorem~\ref{thm-subg} since
  an optimal   $f^*_\ell$ is $\epsilon$-optimal for all $\epsilon>0$.
  \end{proof}

For the next corollary, let $\widehat{\L}(c)$ denote the collection of all loss functions $\ell$ such that 
\[
\ell(y,f^*_\ell(X)) \le g_\ell(y)  \quad \text{a.s.} 
\]
for some function $g_\ell:\Y\to \R_+$ with $\EXP\big[ g_\ell^2(Y)\big] \le c^2$. 

\begin{cor}
  \label{cor2a}
If $T$ is a transformation such that 
\[
 I(Y;X)-I(Y;T(X)) \le \frac{2\delta^2}{c^2},
\]
then  $T$ is universally $\delta$-lossless  for the family $\widehat{\L}(c)$.
\end{cor}

\begin{proof}
Since   $\ell(y,f^*_\ell(X))$ is a.s.\ upper bounded by $g_\ell(y)$
for any $\ell\in \widehat{\L}(c)$, by  Hoeffding's lemma
\cite[Lemma~2.2]{BuLuMa13},  we have that  $\ell(y,f^*_\ell(X)) $ is conditionally  $\frac{g_\ell^2(y)}{4}$-subgaussian given $T(X)$. Thus from Corollary~\ref{cor2}, for all  $\ell\in \widehat{\L}(c)$, we have 
\begin{align*}
L_\ell^*(Y|T(X))- L_\ell^*(Y|X)   &  \le \sqrt{\frac{1}{2}\EXP[g_\ell^2(Y)] \big( I(Y;X)-I(Y;T(X)) \big) } \\
& \le  \sqrt{\frac{c^2}{2} \big( I(Y;X)-I(Y;T(X)) \big) }\\
&\le \delta 
\end{align*} 
if $I(Y;X)-I(Y;T(X)) \le \frac{2\delta^2}{c^2}$. 
\end{proof}

The next corollary generalizes and gives a much simplified  proof of 
Farag\'o and Gy\"orfi \cite{FaGy75}, see also Devroye, Gy\"orfi and Lugosi \cite[Theorem.~32.3]{DeGyLu96}.  This result states  for binary classification ($Y$ is $0$-$1$-valued and $\ell(y,y')=\IND_{y\neq y'}\; $) that if a sequence of functions $T_n:\R^d\to \R^d$ is such that $\|X-T_n(X)\|\to 0$ in probability as $n\to \infty$, then $ L_\ell^*(Y|T_n(X))\to  L_\ell^*(Y|X)$ as $n\to \infty$. 

\begin{cor}
\label{cor3}
Assume that a sequence of transformations  $T_n:\R^d\to \R^d$ is such that   $T_n(X)\to  X$ in distribution  (i.e., $P_{T_n(X)} \to P_X$ weakly) as $n\to \infty$. Then for any bounded loss function $\ell$,
\begin{equation}
    \label{eq_approx}
    \lim_{n\to \infty} L_\ell^*(Y|T_n(X))=  L_\ell^*(Y|X).
\end{equation}
\end{cor}

Note that this corollary and its proof still hold  without any change if  $X$ takes values in an arbitrary complete  separable metric space. 
For example, in the setup of function classification, $X$ may take values in an $L_p$ function space for $1\le p<\infty$, and $T_n$ is a truncated series expansion or a quantizer. 
Interestingly, here the asymptotic losslessness property is guaranteed even in the case where the sequence of transformations $T_n$ and the loss function $\ell$ are not matched at all. 

\begin{proof}
If $T_n(X)\to X$ in distribution, then clearly $(Y,T_n(X))\to (Y,X)$ in distribution. Thus the   lower semicontinuity of mutual information with respect to convergence in distribution 
(see, e.g., Polyanskiy and Wu \cite[Eq.~4.28]{PoWu22})  implies 
\[
\liminf_{n\to \infty} I(Y;T_n(X)) \ge I(Y;X).
\]
Since $ I(Y;T_n(X)) \le I(Y;X)$ for all $n$, we obtain 
\[
\lim_{n\to \infty} I(Y;T_n(X)) = I(Y;X).
\]
Combined with Corollary~\ref{cor1} (with $T$ replaced with $T_n$),  this gives 
\begin{align*}
0\le \lim_{n\to \infty} L_\ell^*(Y|T_n(X))  -L_\ell^*(Y|X) & \le 
\lim_{n\to \infty}   \frac{\|\ell\|_{\infty}}{\sqrt{2}} \sqrt{
  I(Y;X)-I(Y;T_n(X))}   \\
&=0.
\end{align*}
\end{proof}

\section{Applications}

\label{sec-appl}

\subsection{Classification.}
For classification, $\Y$ is the  finite set $\{1,\dots,M\}$ and the cost is the $0-1$ loss
\begin{align*}
\ell(y,y')=\IND_{y\ne y'}.
\end{align*}
In this setup the risk of estimator $f$ is the error probability $\PROB(Y\neq f(X))$.
With the notation
\begin{align*}
P_y(x)=\PROB(Y=y| X=x),
\end{align*}
the optimal estimator is  the Bayes decision
\begin{align*}
f^*(x)=\argmax_{y\in \Y} P_y(x), 
\end{align*}
and the minimum risk is the Bayes error probability
\begin{align*}
L^*(X)=1-\EXP\Big[\max_{y\in \Y} P_y(X)\Big]. 
\end{align*}
If  $L^*(T(X))=1-\EXP\big[\max_y\EXP[P_y(X)\mid T(X)]\big]$ stands for the Bayes error probability for  the transformed observation vector $T(X)$, then (\ref{eq1}) with $\|\ell\|_{\infty}=1$ yields the upper bound 
\[
 L^*(T(X))- L^*(X)   \le  \frac{1}{\sqrt{2}} \sqrt{
  I(Y;X)-I(Y;T(X))};
\]
see also \cite[Corollary~2]{XuRa22} for a similar bound in the context of Bayesian learning. 

As a special case, the feature selector  $X\mapsto X_S$ is lossless if 
\begin{align}
\label{hyp}
L^*(X)=L^*(X_S).
\end{align}
Gy\"orfi and Walk \cite{GyWa17} studied  the corresponding hypothesis testing problem.
Using a $k$-nearest-neighbor ($k$-NN)  estimate of the excess Bayes error probability $L^*(X_S)-L^*(X)$, they introduced a test statistic and accepted the hypothesis (\ref{hyp}), if the test statistic is less than a threshold. Under some mild condition the strong consistency of this test has been proved.

\subsection{Nonparametric regression.}
\label{sec-regr}
For the nonparametric regression problem the cost is the squared loss
\begin{align*}
\ell(y,y')=(y- y')^2, \quad y,y'\in \R, 
\end{align*}
and the best statistical inference is the regression function
\begin{align*}
m(X)=\EXP[Y| X]
\end{align*}
(here we assume $\EXP[Y^2]<\infty$). Then, the minimum risk is the residual variance
\begin{align*}
L_\ell^*(Y|X)=\EXP[(Y-m(X))^2].
\end{align*}

If $L^*(X)=L^*_\ell(Y|X)$ and $L^*(T(X))=L_\ell^*(Y|T(X))$ denote the residual variances for the observation vectors $X$ and $T(X)$, respectively, then 
\begin{align*}
L^*(T(X))-L^*(X)
&=\EXP\big[ (Y-\EXP[m(X)\mid T(X)] )^2\big] -\EXP\big[ (Y-m(X))^2\big] \\
&=\EXP\big[ (m(X)-\EXP[m(X)\mid T(X)])^2\big].
\end{align*}
Note that the excess residual variance $L^*(T(X))-L^*(X)$ does not depend on the distribution of the residual $Y-m(X)$.

Next we show that the conditions of Corollary~\ref{cor2} hold with $f^*_\ell(x)=m(x)$ for the important  case 
\[Y=m(X)+N,
\]
where $N$ is a zero-mean noise variable that is independent of $X$ and satisfies $\EXP[N^4]<\infty$, and  $m$ is bounded as   $|m(x)|\le K$ for all $x$. For this model  we have 
\[
\ell(y,f^*(X))
= (y-m(X))^2 
\le (|y|+|m(X)|)^2 
\le (|y|+K)^2. 
\]
Thus $\ell(y,f^*(X))$ is a nonnegative random variable a.s.\ bounded by $(|y|+K)^2$, which implies via  Hoeffding's lemma (e.g.,  \cite[Lemma~2.2]{BuLuMa13})  that it is conditionally 
$\sigma^2(y)$-subgaussian given $T(X)$ with $\sigma^2(y)= \frac{(|y|+K)^4}{4}$. We have 
\begin{align*}
\EXP[\sigma^2(Y)] 
&= \frac{\EXP\left[(|Y|+K)^4\right]}{4} \\ 
&\le \frac{\EXP\left[(|N|+|m(X)|+K)^4\right]}{4}\\
&\le \frac{\EXP\left[(|N|+2K)^4\right]}{4}\\
&\le \frac{8\EXP\left[|N|^4+(2K)^4\right]}{4}\\
&\le 2\EXP[N^4] +32K^4\\
&<\infty, 
\end{align*}
so the conditions of  Corollary~\ref{cor2} hold and we obtain 
\begin{align*}
L_\ell^*(Y|T(X))- L_\ell^*(Y|X)  
&= 
\EXP[(Y-\EXP[Y|T(X)])^2] - \EXP[(Y-\EXP[Y|X])^2] \\
& \le \sqrt{ \big(2\EXP[N^4] +32K^4\big)  \big( I(Y;X)-I(Y;T(X)) \big) }\, .
\end{align*}

Again, the feature selection $X_S$ is called lossless, when $L^*(X)=L^*(X_S)$ holds.
As a test statistic,
Devroye et al.\ \cite{DeGyLuWa18} introduced a 1-NN  estimate of $L^*(X_S)-L^*(X)$
and proved the strong consistency of the corresponding test.

\subsection{Portfolio selection.}
\label{sec-portf}

The next example is related to  the negative of the log-loss or log-utility; see
Algoet and Cover \cite{AlCo88},
Barron and Cover \cite{BaCo88},
Chapters 6 and 16 in Cover and Thomas \cite{CoTh06},
Gy\"orfi et al.\ \cite{GyOtUr11}.

Consider a market consisting of $d_a$ assets.
The evolution of the market in time is represented by a
sequence of (random) price vectors $S_1,S_2,\ldots \in \R_+^{d_a}$ with 
\[
S_n=(S_n^{(1)},\dots ,S_n^{(d_a)}), 
\]
where 
the $j$th component $S_n^{(j)}$ of $S_n$ denotes
the price of the $j$th asset
in the $n$th trading period.
Let us transform the sequence of price vectors $\{S_n\}$ into the sequence of
return (relative price) vectors $\{R_n\}$ defined as 
\[
R_n=(R_n^{(1)},\dots ,R_n^{(d_a)}),
\]
where 
\[
R_n^{(j)}=\frac{S_n^{(j)}}{S_{n-1}^{(j)}}.
\]

The constantly rebalanced portfolio selection
is a multi-period investment strategy, where at the beginning of
each trading period the investor redistributes the wealth among the assets.
The investor is allowed to
diversify their capital at the beginning of each trading period
according to a portfolio vector $b=(b^{(1)},\dots b^{(d_a)})$. The
$j$th component $b^{(j)}$ of $b$ denotes the proportion of the
investor's capital invested in asset $j$. Here we  assume that the portfolio vector $b$ has nonnegative components with $\sum_{j=1}^{d_a} b^{(j)} =1$.  The simplex of possible portfolio
vectors is denoted by $\Delta_{d_a}$.

Let $S_0=1$ denote the investor's initial capital.
Then at the beginning of the first trading period
$S_0 b^{(j)}$ is invested into asset $j$,
and it results in return $S_0b^{(j)}R_1^{(j)}$, and therefore
at the end of the first trading period
the investor's  wealth becomes
\[
S_1 = S_0 \sum_{j=1}^{d_a} b^{(j)}R_1^{(j)} = \ip{b}{R_1},
\]
where $\ip{\,\cdot\,}{\,\cdot\,}$ denotes the standard inner product in $\R^{d_a}$.
For the second trading period, $S_1$ is the new initial capital
\[
S_2=S_1 \cdot \ip{b}{R_2}
= \ip{b}{R_1} \cdot \ip{b}{R_2}.
\]
By induction, for the trading period $n$ the initial capital
is $S_{n-1}$, and therefore
\[
S_n =  S_{n-1} \ip{b}{R_n}=\prod_{i=1}^n \ip{b}{R_i}.
\]
The asymptotic average growth rate of this portfolio selection strategy is
\begin{align*}
\lim_{n\to \infty} \frac 1n \log S_n
&=\lim_{n\to \infty} \frac 1n \sum_{i=1}^n \log\ip{b}{R_i}
\end{align*}
assuming the limit exists. 

If the market process $\{R_i\}$ is memoryless, i.e.,
it is a sequence of i.i.d.\ random return vectors, 
then the strong law of large numbers implies that the best constantly rebalanced portfolio (BCRP) is
the log-optimal portfolio:
\[
b^*=\argmax_{b\in \Delta_{d_a}} \EXP\big[ \log\ip{b}{R_1} \big],
\]
while the best asymptotic average growth rate is
\[
W^*=\max_{b\in \Delta_{d_a}} \EXP\big[ \log\ip{b}{R_1}\big].
\]

Barron and Cover \cite{BaCo88} extended this setup to portfolio selection with side information.
Assume that $X_1,X_2,\dots$ are $\R^d$ valued side information vectors such that $(R_1,X_1), (R_2,X_2),\ldots$ are i.i.d.\ and in each round $n$ the portfolio vector may depend on $X_n$.
The strong law of large numbers yields
\begin{align*}
\lim_{n\to \infty} \frac 1n \log S_n
&=\lim_{n\to \infty} \frac 1n \sum_{i=1}^n \log\ip{b(X_i)}{R_i}
= \EXP\big[ \log\ip{b(X_1)}{R_1}\big]  \quad \text{a.s.}
\end{align*}
Therefore, the log-optimal portfolio has the form
\[
b^*(X_1)=\argmax_{b\in \Delta_{d_a}} \EXP\big[ \log\ip{b}{R_1}\mid X_1\big] 
\]
and the best asymptotic average growth rate is
\[
W^*(X)=\EXP\left[\max_{b\in \Delta_{d_a}} \EXP\big[ \log\ip{b}{R_1}\mid X_1\big] \right].
\]

Barron and Cover \cite[Thm.~2]{BaCo88} proved that
\begin{equation}
\label{BaCoibound} 
W^*(X)-W^*\le I(R_1;X_1).  
\end{equation}
The next theorem generalizes this result by upper bounding the loss of the best asymptotic growth rate when instead of $X$, only degraded side information $T(X)$ is available.

\begin{theorem}
  \label{BaCoThm}
For any measurable $T: \R^d \to \R^{d'}$,
\[
    W^*(X)-W^*(T(X))\le I(R_1;X_1)-I(R_1;T(X_1))
\]
 assuming the terms on the right hand side are finite. 
\end{theorem}

\noindent \emph{Remarks:} 
\vspace{-3pt}

\begin{enumerate}[label=(\roman*)] 

\item  
As in Theorem~\ref{thm-subg}, the difference $I(R_1;X_1)-I(R_1;T(X_1))$ in the upper bound is equal to  $I(R_1;X_1|T(X_1))$, a quantity that is always nonnegative but may be equal to $\infty$. In this case, we interpret the right hand side as $\infty$. 
\item
There is a correspondence between this setup of portfolio selection and the setup in previous sections. In particular,  $Y$ from the previous sections is equal to $R$ with range $\R_+^{d_a}$ and the inference is $b(X)$ taking values in $\Delta_{d_a}$.
Then, the loss is $-\log\ip{b(X)}{R}$.
If we assume that for all $j=1,\dots d_a$, 
\begin{align}
\label{c*} 
|\log R^{(j)}|\le c_{max} \quad \text{a.s.}, 
\end{align}
then
\begin{align*}
|\log\ip{b(X)}{R}|\le c_{max} \quad \text{a.s.}
\end{align*}
and so Corollary  \ref{cor1} implies
\[
    W^*(X)-W^*(T(X))\le  \frac{c_{max}}{\sqrt{2}}\sqrt{I(R_1;X_1)-I(R_1;T(X_1))}.
\]
Note that from the point of view of applications, (\ref{c*}) is a mild condition. 
For example, for NYSE daily data $c_{max}\le 0.3$;  see  
Gy\"orfi et al.\  \cite{GyOtWa17}.
\end{enumerate}

\begin{proof}
  Let $(R,X)$ be a generic copy of the $(R_i,X_i) $ . 
  Writing out explicitly the dependence of $W^*$ on $P_{\, R}$, we
  have
  \[
  W^*(X) = \int W^*(P_{\, R|X=x}) P_X(dx)
\]
and from \eqref{idefalt} we have 
\[
I(R; X)= \int D(P_{\, R|X=x} \|P_{\, R})  P_X(dx).
\]
Thus the bound $W^*(X)-W^*\le I(R_1;X_1)$ in \eqref{BaCoibound} can be written as 
 \begin{align}
 \int  W^*(P_{\, R|X=x})  P_X(dx) - W^*  &\le  \int D(P_{\, R|X=x} \|P_{\, R})
             P_X(dx).\label{eqdbound} 
 \end{align}
Furthermore, letting $Z=T(X)$, we  have 
 \begin{align*}
    W^*(X) - W^*(Z)  &= \int  W^*(P_{\, R|X=x}) P_X(dx) -  \int
                       W^*(P_{\, R|Z=z}) P_Z(dz). 
 \end{align*}
Since $R\to X\to Z$ is a Markov chain, $P_{\, R|X=x} = P_{\,
  R|X=x,Z=z}$, and we obtain
\begin{align}
  \MoveEqLeft  W^*(X) - W^*(Z)\nonumber \\
  &=  \int \Bigg( \int  W^*(P_{\,R|X=x,Z=z})    P_{X|Z=z}(dx)  -
    W^*(P_{\, R|Z=z}) \Bigg) P_Z(dz) . \nonumber 
\end{align}
Applying \eqref{eqdbound} with  $W^*(P_{\, R|X=x})$ replaced with $W^*(P_{\,R|X=x,Z=z})$  and $ W^*$ replaced with $W^*(P_{\, R|Z=z})$ with $z$ fixed,
we can bound the expression in parentheses as 
\begin{align*}
   \MoveEqLeft[8]   \int  W^*(P_{\,R|X=x,Z=z})    P_{X|Z=z}(dx)  -
    W^*(P_{\, R|Z=z})  \\
    & \le  \int D(P_{\,R|X=x,Z=z}\| P_{\, R|Z=z})
    P_{X|Z=z}(dx), 
\end{align*}
and therefore  
\begin{align}
  \MoveEqLeft W^*(X) - W^*(Z)\nonumber \\
  & \le  \int   \int D(P_{\,R|X=x,Z=z}\| P_{\, R|Z=z})
    P_{X|Z=z}(dx) P_Z(dz) \nonumber \\
  & = I(R;X|Z),  \label{condzbound} 
\end{align}
where \eqref{condzbound} follows from the alternative
expression   \eqref{cond-inf-def1} of the conditional mutual
information.

As in the proof Theorem~\ref{thm-subg}, the conditional independence of
$R$ and $Z=T(X)$ given $X$ implies
\[
  I(R;X|Z)=  I(R;X)-I(R;T(X)),
\]
which completes the proof. 
\end{proof}

\subsection{Information bottleneck.}
\label{sec-ib} 
Let $X$ and $Y$ be random variables as in Section~\ref{sec-lossless}. When $Y\to X\to Z$,  the joint distribution $P_{YXZ}$ of the triple $(Y,X,Z)$ is determined (for fixed $P_{YX}$) by the conditional distribution (transition kernel)  $P_{Z|X}$ as $P_{YXZ}=P_{YX}P_{Z|X}$. The information bottleneck (IB) framework can be formulated as 
the study of the  constrained optimization problem
\begin{equation}
\label{eq-ib}
\begin{array}{cl}
  \text{maximize \ \ \ }   &  I(Y;Z)  \\
             &  \vspace{-10pt} \\
    \text{subject to\ \ \ }  &  I(X;Z)  \le  \alpha 
\end{array}
\end{equation}
for a given $\alpha> 0$, where the maximization  is over all transition kernels $P_{Z|X}$. 

Originally proposed  by Tishby et al.\  \cite{TiPeBi99}, the solution of the IB problem is a transition kernel $P_{Z|X}$, interpreted as a stochastic transformation,  that ``encodes" $X$ into a ``compressed" representation $Z$ that preserves relevant information about  $Y$ through maximizing $ I(Y;Z)$, while  compressing $X$ by requiring that $I(X;Z)\le \alpha$. The intuition behind this framework is that by maximizing $I(Y;Z)$, the representation $Z$ will retain the predictive power of $X$  with respect to $Y$, while the requirement $ I(X;Z) \le \alpha $ makes the representation $Z$ concise. 

Note that in case $X$ is discrete and has finite entropy $H(X)$,  setting $\alpha = H(X)$, or setting formally $\alpha=\infty$ in the general case, the constraint $ I(X;Z)  \le  \alpha $ becomes vacuous and (assuming the alphabet of $Z$ is sufficiently large) the resulting $Z$ will achieve the  upper bound  $I(Y;Z)= I(Y;X)$, so that  $I(Y;X|Z)=I(Y;X)-I(Y;Z)=0$, i.e., $Y\to Z\to X$. Thus the solution to \eqref{eq-ib} can be considered as a stochastically relaxed version of a minimal sufficient statistic for $X$ in predicting $Y$ (see  Goldfeld and Polyanskiy \cite[Section~II.C]{GoPo20} for more on this interpretation). Recent tutorials on the IB problem include  Asoodeh and  Calmon \cite{AsCa20} and Zaidi et al.\ \cite{Zai-etal20}. 

Theorem~\ref{thm-subg} and its corollaries can be used to motivate the IB principle from an estimation-theoretic viewpoint. Let 
\[
I(\alpha) = \sup_{P_{Z|X}: I(X;Z)\le \alpha} I(Y;Z)
\]
be the  value function for \eqref{eq-ib} and $Z_{\alpha}$ a resulting optimal $Z$ (assuming such a maximizer exists). From the remark after Theorem~\ref{thm-subg} we know that the bounds given in the theorem and in its corollaries remain valid if we replace $T(X)$ with a random variable $Z$ such that $Y\to X\to Z$. Then, for example, Corollary~\ref{cor1}  implies that 
\[
 L_\ell^*(Y|Z_{\alpha}) - L_\ell^*(Y|X)   \le  c \sqrt{ I(Y;X)- I(\alpha)}
\]
for all $\ell$ such that $\|\ell\|_{\infty}\le \sqrt{2} c$. 

Thus the IB paradigm minimizes,  under the complexity constraint $I(X;Z)$ $\le$ $\alpha$,  an upper bound on the difference $L_\ell^*(Y|Z) - L_\ell^*(Y|X)  $  
that \emph{universally holds} for all loss functions $\ell$ with $\|\ell\|_{\infty}\le \sqrt{2} c$. The resulting $Z_{\alpha}$  will then  have \emph{guaranteed performance} in predicting $Y$ with respect to \emph{all} sufficiently bounded loss functions. This gives an  novel operational interpretation of the IB framework that seems to have been overlooked in the literature.

\subsection{Deep learning} 
\label{sec-dl} 
The IB paradigm can also serve as a learning objective in deep neural networks (DNNs).  
Here  the Lagrangian relaxation of \eqref{eq-ib} is considered.
In particular, letting $X$ denote the input and $Z^{\theta}$ the output of the last hidden layer of the DNN, where $\theta \in \Theta\subset \R^K$ is the collection of network parameters (weights), the objective is to maximize 
\begin{equation}
\label{eq-ib-lgr}
    I(Y;Z^{\theta}) -\beta I(X;Z^{\theta}) 
\end{equation}
over $\theta\in \Theta$ for a given $\beta>0$.  The parameter $\beta$ controls the trade-off between how informative $Z^{\theta}$ is about $Y$, measured by $I(Y;Z^{\theta})$,  and how much $Z^{\theta}$ is ``compressed,"  measured by $I(X;Z^{\theta})$.  Clearly, larger values of $\beta$ correspond to smaller values of $I(X;Z^{\theta})$ and thus to more compression. Here $Z^{\theta}$ is  either  a deterministic function of $X$ in the form of $Z^{\theta}=T^{\theta}(X)$, where 
$T^{\theta}:\R^d \to \R^{d'}$ represents the deterministic DNN, or it is produced by a stochastic kernel $P^{\theta}_{Z|X}$, parametrized by the network parameters $\theta\in \Theta$. This latter is achieved by injecting independent noise into the network's intermediate layers. 

In addition to the motivation  explained in the previous section, the IB framework for  DNNs can be thought as a regularization method that results in improved generalization capabilities for a network trained on data via stochastic gradient based methods, see, e.g.,  
Tishby and Zaslavsky \cite{TiZa15}, 
Shwartz-Ziv and Tishby \cite{ScTi17}, 
Alemi et al.~\cite{AlFiDiMu17}, as well as many other references in the excellent survey article 
Goldfeld and Polyanskiy \cite{GoPo20}, and the special issue \cite{GeKu20} on information bottleneck and deep learning. 

 As in the previous section, our Theorem~1 and corollaries can serve as a (partial) justification for setting \eqref{eq-ib-lgr} as a learning objective. Assume that after training with a given $\beta>0$,  the obtained $Z^{\theta(\beta)}$ has (true)  mutual information $I(Y;Z^{\theta(\beta)})$ with $Y$ (typically,  this will not be the optimal solution since maximizing \eqref{eq-ib-lgr} is not feasible and in practice only a proxy lower bound is optimized during training, see, e.g.,  Alemi et al.~\cite{AlFiDiMu17}). Then by Corollary~\ref{cor1} the obtained network has guaranteed predictive performance 
 \[
 L_\ell^*(Y|Z^{\theta(\beta)}) \le   L_\ell^*(Y|X) +  c\epsilon  
\]
for all loss functions $\ell$ with   $\|\ell\|_{\infty}\le \sqrt{2} c$, where 
\[
\epsilon = \sqrt{ I(Y;X)- I(Y;Z^{\theta(\beta)})}\;. 
\]

\section{Proof of Theorem~\ref{coro0}}

\label{sec-proof}

\begin{proof}[Proof of (a)] 
The bounds given in the proof of Theorem~1 in  \cite{GyWa12}  imply
\begin{align*}
L_n
&\le
J_{n,1}+J_{n,2}+J_{n,3}+J_{n,4}+J_{n,5},
\end{align*}
where
\begin{spreadlines}{2ex} 
\begin{align*}
J_{n,1}
&=
\sum_{A\in \P_n,B\in \Q_n,C\in {\cal R}_n}
\left|P^n_{XYZ}(A,B,C)
-P_{XYZ}(A,B,C)
\right|  , \\
J_{n,2}
&=
\sum_{B\in \Q_n,C\in {\cal R}_n}
\left|P_{YZ}(B,C)
-P^n_{YZ}(B,C)
\right| ,   \\
J_{n,3} &=
\sum_{A\in \P_n,C\in {\cal R}_n}
\left|P_{XZ}(A,C)-P^n_{XZ}(A,C)\right|, \\
J_{n,4}
&=
\sum_{C\in {\cal R}_n} \left|P^n_{Z}(C)-P_Z(C)\right| , \\
\intertext{and} 
J_{n,5}
&=
\sum_{A\in \P_n,B\in \Q_n,C\in {\cal R}_n}
\left|P_{XYZ}(A,B,C)
-\frac{P_{XZ}(A,C)P_{YZ}(B,C)}{P_{Z}(C)}
\right| \, .
\end{align*}
\end{spreadlines}

Using  large deviation inequalities from  Beirlant at al. \cite{BeDeGyVa01}
and in Biau and Gy\"orfi \cite{BiGy05},
Gy\"orfi and Walk \cite{GyWa12} proved that
 for all $\varepsilon_i>0$, $i=1,\dots ,4$ and $\delta>0$,
\begin{align}
\MoveEqLeft\PROB (L_n>\varepsilon_1+\varepsilon_2+\varepsilon_3+\varepsilon_4+\delta )\nonumber\\
&\le
\PROB (J_{n,1}>\varepsilon_1 )
+\PROB (J_{n,2}>\varepsilon_2 )
+\PROB (J_{n,3}>\varepsilon_3 )
+\PROB (J_{n,4}>\varepsilon_4 )
+\IND_{J_{n,5}>\delta}\nonumber\\
&\le
2^{m_n\cdot m'_n\cdot m''_n}e^{-n\varepsilon_1^2/2}
+2^{m'_n\cdot m''_n}e^{-n\varepsilon_2^2/2}
+2^{m_n\cdot m''_n}e^{-n\varepsilon_3^2/2}+2^{m''_n}e^{-n\varepsilon_4^2/2}\nonumber\\
&\quad +\IND_{ J_{n,5} >\delta}.
\label{*}
\end{align}
We note that  bounds on the probabilities $\PROB (J_{n,i}>\varepsilon_i )$ for $i=1,\ldots,4$ were proved   in  \cite{GyWa12} without either assuming the null hypothesis $\mathcal H_0$ or using the condition that the partitions are nested. 
Under the null hypothesis, Gy\"orfi and Walk \cite{GyWa12} claimed that
\begin{align*}
J_{n,5} =0.
\end{align*}
As Neykov et al.\ \cite{NeBaWa21} observed, this was incorrect. In order to resolve the gap, we show that under $\lim_n h_n=0$ and condition \eqref{Lip} and under the null hypothesis, the last term in (\ref{*}) is $o(1)$, i.e.,
\begin{align*}
\IND_{\sum_{A\in \P_n,B\in \Q_n,C\in {\cal R}_n}
\left|P_{XYZ}(A,B,C)-\frac{P_{XZ}(A,C)P_{YZ}(B,C)}{P_{Z}(C)}\right|   >\delta}=0
\end{align*}
if $n$ is large enough.
The null hypothesis implies that
\begin{align*}
\MoveEqLeft[6] P_{XYZ}(A,B,C)-\frac{P_{XZ}(A,C)P_{YZ}(B,C)}{P_{Z}(C)}\\
&=
\PROB(X\in A, Y\in B, Z\in C)-\frac{P_{XZ}(A,C)\PROB(Y\in B, Z\in C)}{P_{Z}(C)}\\
&=
\EXP\big[ \PROB(X\in A, Y\in B\mid Z)\IND_{Z\in C}\big] 
 -\EXP\big[\PROB(Y\in B\mid Z)\IND_{Z\in C}\big]\frac{P_{XZ}(A,C)  }{P_{Z}(C)}\\
&=
\EXP\big[ \PROB(X\in A\mid Z) \PROB(Y\in B\mid Z)\IND_{Z\in C}\big] 
-\EXP\big[\PROB(Y\in B\mid Z)\IND_{Z\in C}\big]\frac{P_{XZ}(A,C)  }{P_{Z}(C)}.
\end{align*}
Thus,
\begin{align*}
\MoveEqLeft[4]
\sum_{A\in \P_n,B\in \Q_n,C\in {\cal R}_n}
\left|P_{XYZ}(A,B,C)-\frac{P_{XZ}(A,C)P_{YZ}(B,C)}{P_{Z}(C)}\right|\nonumber\\
&\le
\sum_{A\in \P_n,B\in \Q_n,C\in {\cal R}_n}
\EXP\left[\PROB(Y\in B\mid Z)\left|\PROB(X\in A\mid Z) \IND_{Z\in C}
-\IND_{Z\in C}\frac{P_{XZ}(A,C)  }{P_{Z}(C)}\right|\right]\nonumber\\
&=
\sum_{A\in \P_n,C\in {\cal R}_n}
\EXP\left[\left|\PROB(X\in A\mid Z) \IND_{Z\in C}
-\IND_{Z\in C}\frac{P_{XZ}(A,C)  }{P_{Z}(C)}\right|\right].
\end{align*}
Let $p(\,\cdot\mid z)$ and $C_n(z)$ be as in the Condition \ref{cond1}.
Then,
\begin{align}
\MoveEqLeft[4]
\sum_{A\in \P_n,C\in {\cal R}_n}
\EXP\left[\left|\PROB(X\in A\mid Z) \IND_{Z\in C}
-\IND_{Z\in C}\frac{P_{XZ}(A,C)  }{P_{Z}(C)}\right|\right]\nonumber\\
&=
\sum_{A\in \P_n,C\in {\cal R}_n}
\int_C\left|\int_Ap(x\mid z)P_X(dx) 
-\frac{\int_A[\int_Cp(x\mid z')P_Z(dz') ]P_X(dx)   }{P_{Z}(C)}\right|P_Z(dz)\nonumber\\
&\le 
\int\int\left|p(x\mid z)
-\frac{\int_{C_n(z)}p(x\mid z')P_Z(dz')    }{P_{Z}(C_n(z))}\right|P_Z(dz)P_X(dx)\nonumber\\
&\le
C^*h_n,
\label{**}
\end{align}
where in the last step we used  condition (\ref{Lip}). 
The inequalities (\ref{*}) and (\ref{**})    imply that
\begin{align*}
\MoveEqLeft \PROB\bigg(L_n
>
c_1\Big(\sqrt{\frac{m_nm'_nm''_n}{n}} + \sqrt{\frac{m'_nm''_n}{n}}
+ \sqrt{\frac{m_nm''_n}{n}}+\sqrt{\frac{m''_n}{n}}\,\Big) +(\log n)h_n \bigg)\\
&\le
4e^{-(c_1^2/2-\log 2)m''_n}+\IND_{ C^* h_n  >(\log n) h_n}\\
&\le
4e^{-(c_1^2/2-\log 2)m''_n},
\end{align*}
if $n\ge e^{C^*}$. Since $m_n''$ is proportional to $1/h_n^{d'}$, condition \eqref{hnS2} on $h_n$ implies $\sum_{n=1}^{\infty} \PROB(L_n \ge t_n)  <\infty$,  and thus by the  Borel-Cantelli lemma,  after a random sample size, the test makes  no error with probability one.  
\end{proof}

\begin{proof}[Proof of (b)] 
The proof  is a refinement of  the proof of Corollary 1 in \cite{GyWa12} in which  we avoid the condition there that the sequences of partitions $\P_n$ and $\Q_n$ are nested.
According to the proof of Part (a) (see the remark after \eqref{*}), we get that
\[
\liminf_{n\to\infty}L_n
\ge 
\liminf_{n\to\infty}(L_n-J_{n,5})
+
\liminf_{n\to\infty}J_{n,5}
=
\liminf_{n\to\infty}J_{n,5} \quad \text{a.s.} 
\]
To simplify the notation, let $P_{XY|z}=P_{XY|Z=z}$, $P_{X|z}=P_{X|Z=z}$,  and $P_{Y|z}=P_{Y|Z=z}$. 
Let $L^*$ be the expected total variation distance between  $P_{XY|z}$ and $P_{X|z} P_{Y|z}\,$:
\[
L^*=  \int \sup_F \big| P_{XY|z} (F)-P_{X|z}P_{Y|z} (F)\big| \, P_Z(dz),
\]
where the supremum is taken over all Borel subsets $F$ of $\mathbb R^d\times \mathbb R^{d'}$.
It suffices to prove that by the condition $\lim_n h_n=0$,
\[
\liminf_{n\to\infty} J_{n,5}
\ge
2L^*>0.
\]
One has that
\begin{align*}
\MoveEqLeft[6] \sum_{{A\in \P_n,B\in \Q_n,C\in {\cal R}_n}}
\left|P_{XYZ}(A,B,C)
-\frac{P_{XZ}(A,C)P_{YZ}(B,C)}{P_{Z}(C)}
\right|\\
&\ge
\int
\sum_{A\in \P_n,B\in \Q_n}\left|P_{XY|z}(A,B)
-P_{X|z}(A)P_{Y|z}(B)
\right|P_Z(dz)-W_n,
\end{align*}
where
\begin{align}
\!\!\!\! W_n
&\le
\sum_{A\in \P_n,B\in \Q_n}\int
\left|\frac{\int_{C_n(z)}P_{XY|z'}(A,B)P_Z(z')}{P_Z(C_n(z))}
-
P_{XY|z}(A,B)\right|P_Z(dz) \label{eq-ft} \\
& \quad + 
\sum_{A\in \P_n}\int
\left|\frac{\int_{C_n(z)}P_{X|z'}(A)P_Z(dz')}{P_Z(C_n(z))}
-P_{X|z}(A)
\right|P_Z(dz)  \label{eq-st} \\
 &\quad   +
\sum_{B\in \Q_n}\int
\left|\frac{\int_{C_n(z)}P_{Y|z'}(B)P_Z(dz')}{P_Z(C_n(z))}
-P_{Y|z}(B)
\right| P_Z(dz)  \label{eq-tt}
\end{align}
In  \cite{GyWa12} it was shown that the condition $\lim_n h_n=0$  implies $\lim_{n} W_n= 0$ if the sequence of partitions $\{\P_n,Q_n\}_{n\ge 1}$ is nested. 
In order to avoid this nestedness condition, introduce the density $p(x,y|z)$ of the conditional distribution $P_{XY|z}$ with respect to the distribution $P_{XY}$ of $(X,Y)$ as a dominating measure, 
and similarly let $p_n(x,y|z)$ be the density of the conditional distribution $\int_{C_n(z)}P_{XY|z'}(\,\cdot\,,\,\cdot\,)P_Z(dz')/P_Z(C_n(z))$ with respect to $P_{XY}$,
i.e., $p_n(x,y|z)=\int_{C_n(z)}p(x,y|z')P_Z(dz')/P_{Z}(C_n(z))$.
Then,
\begin{align*}
\MoveEqLeft[8] \sum_{A\in \P_n,B\in \Q_n}
\left|\frac{\int_{C_n(z)}P_{XY|z'}(A,B)P_Z(dz')}{P_Z(C_n(z))}
-
P_{XY|z}(A,B)\right|\\
&\le
\int\int |p_n(x,y|z)-p(x,y|z)|P_{XY}(dx,dy),
\end{align*}
and therefore the  term on  the right hand side of \eqref{eq-ft}  will converge to zero as long as   
\begin{align*}
\int\int\int |p_n(x,y|z)-p(x,y| z)|P_{XY}(dx,dy)P_Z(dz)\to 0,
\end{align*}
which follows from $\lim_n h_n=0$ using the standard technique of the bias of partitioning regression estimate for the regression function $p(\, \cdot \, ,\, \cdot \, | z)$;  see Theorem 4.2 in  \cite{GyKoKrWa02}.
The terms in \eqref{eq-st}  and \eqref{eq-tt} can be dealt with analogously.
Thus,
\begin{align*}
\liminf_{n\to\infty} J_{n,5}
&\ge
\liminf_{n\to\infty}
\int
\sum_{A\in \P_n,B\in \Q_n}\left|P_{XY|z}(A,B)
-P_{X|z}(A)P_{Y|z}(B)
\right| P_Z(dz) .
\end{align*}
For fixed $z$, $\lim_n h_n=0$   implies 
\begin{align*}
\MoveEqLeft[6] \lim_{n\to\infty}
\sum_{A\in \P_n,B\in \Q_n}\left|P_{XY|z} (A,B)
-P_{X|z}(A)P_{Y|z}(B) \right|\\
&=
2 \sup_F \big| P_{XY|z} (F)-P_{X|z}P_{Y|z} (F)\big|,
\end{align*}
see Abou-Jaoude \cite{Abo76} and  Csisz\'ar \cite{Csi73}.
Therefore, the dominated convergence theorem yields
\begin{align*}
\MoveEqLeft[5]
\lim_{n\to\infty}
\int \!\!\!
\sum_{A\in \P_n,B\in \Q_n}\left|P_{XY|z} (A,B)
-P_{X|z}(A)P_{Y|z}(B) \right|P_Z(dz)\\
&=
2\int \sup_F \big| P_{XY|z} (F)-P_{X|z} P_{Y|z} (F)\big| \, P_Z(dz) \\
&=
2L^*.
\end{align*}

\end{proof} 

Note that  in the proof of Part (b) the condition (\ref{Lip}) is not used, at all.

\section{Concluding remarks}

\label{sec-concl}

We studied the excess minimum risk in statistical inference and under mild conditions gave a strongly consistent procedure to test from data if a given transformation of the observed feature vector results in zero excess minimum risk for all loss functions. 
It is an open research problem whether  a strong universal test exists,  i.e., a test that is strongly consistent without any condition on the transformation and on the underlying distribution.
We also  developed information-theoretic upper bounds on the excess risk that uniformly hold over  fairly general classes of loss functions. 
The bounds have not been stated in their possible most general form in that the observed quantities were restricted to take values in Euclidean spaces and we did not allow  transformations that are random functions of the observation, both of which restrictions can be relaxed. 
The bounds are possible to sharpen, e.g., in specific cases, but in their  present form are already useful. For example, they give additional theoretical motivation for applying the information bottleneck approach in deep learning.

\end{document}